\documentclass[superscriptaddress,twocolumn,notitlepage]{revtex4-1}

\usepackage[]{amsmath}  			
\usepackage{amssymb}    			
\usepackage{array}      			
\usepackage{comment}
\usepackage{color}
\usepackage{xcolor}

\usepackage{graphicx}           	
\usepackage[update]{epstopdf}   	

\makeatletter
\DeclareOldFontCommand{\rm}{\normalfont\rmfamily}{\mathrm}
\DeclareOldFontCommand{\sf}{\normalfont\sffamily}{\mathsf}
\DeclareOldFontCommand{\tt}{\normalfont\ttfamily}{\mathtt}
\DeclareOldFontCommand{\bf}{\normalfont\bfseries}{\mathbf}
\DeclareOldFontCommand{\it}{\normalfont\itshape}{\mathit}
\DeclareOldFontCommand{\sl}{\normalfont\slshape}{\@nomath\sl}
\DeclareOldFontCommand{\sc}{\normalfont\scshape}{\@nomath\sc}
\makeatother

\begin{document}

\title{Kirigami Actuators}

\author{Marcelo A. Dias}
\email{madias@eng.au.dk}
\affiliation{Department of Engineering, Aarhus University, Inge Lehmanns Gade 10, 8000 Aarhus C, Denmark.}
\affiliation{Department of Physics and Astronomy, James Madison University, Harrisonburg, VA 22807, USA.}

\author{Michael P. McCarron}
\affiliation{
Department of Mechanical Engineering, Boston University, Boston, MA, 02215, USA.
}

\author{Daniel Rayneau-Kirkhope}
\affiliation{
Department of Applied Physics, Aalto University, FI-02150 Espoo, Finland.
}

\author{Paul Z. Hanakata}
\affiliation{
Department of Physics, Boston University, Boston, MA 02215, USA.
}

\author{David K. Campbell}
\affiliation{
Department of Physics, Boston University, Boston, MA 02215, USA.
}

\author{Harold S. Park}
\affiliation{
Department of Mechanical Engineering, Boston University, Boston, MA, 02215, USA.
}

\author{Douglas P. Holmes}
\email{dpholmes@bu.edu}
\affiliation{
Department of Mechanical Engineering, Boston University, Boston, MA, 02215, USA.
}

\date{\today}

\makeatother

\begin{abstract}
Thin elastic sheets bend easily and, if they are patterned with cuts, can deform in sophisticated ways. Here we show that carefully tuning the location and arrangement of cuts within thin sheets enables the design of mechanical actuators that scale down to atomically--thin 2D materials. We first show that by understanding the mechanics of a single, non--propagating crack in a sheet we can generate four fundamental forms of linear actuation: roll, pitch, yaw, and lift.  Our analytical model shows that these deformations are only weakly dependent on thickness, which we confirm with experiments at centimeter scale objects and molecular dynamics simulations of graphene and MoS$_{2}$ nanoscale sheets. We show how the interactions between non--propagating cracks can enable either lift or rotation, and we use a combination of experiments, theory, continuum computational analysis, and molecular dynamics simulations to provide mechanistic insights into the geometric and topological design of kirigami actuators.
\end{abstract}

\maketitle

\begin{figure}[h!]
\begin{center}
\includegraphics[width=1.0\columnwidth]{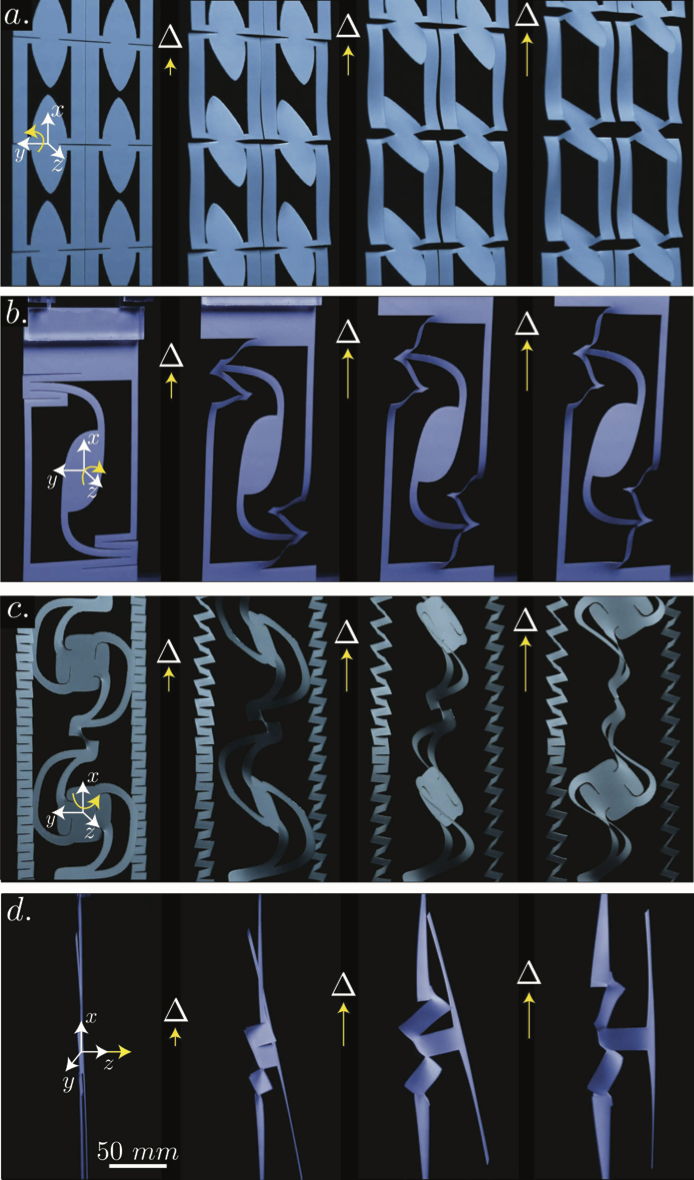}
\end{center}
\vspace{-5mm}
\caption{\footnotesize Examples of linear actuators from kirigami cut patterns. Extension, or applied displacement $\Delta$, along the $x$-direction causes $a.$ rotation about $y$--axis or pitch, $b.$ rotation about $z$--axis or yaw, $c.$ rotation about $x$--axis or roll, and $d.$ out--of--plane deflection in the $z$-direction.\label{fig1}}
\end{figure}

Deformations that bend a material without stretching involve a very low amount of stored elastic energy, and therefore present an opportunity to enable morphing at minimal energetic cost. The potential to exploit these energetically favorable and soft modes has recently emerged with kirigami-based thin sheets~\cite{Reis2015,Seffen2016,Rafsanjani2017,Bico2017}, in which the introduction of cuts has been utilized to give unique structural properties and non-linear behavior, such as auxeticity~\cite{Virk2013,Scarpa2013,Tang2017}, significantly enhanced stretchability~\cite{Shyu2015,Vachicouras2017}, flexible electronic devices~\cite{Zhang2015}, and topologically guided morphings~\cite{Sadoc2012, Charvolin2011, Sadoc2013, Castle2014, Sussman2015, Chen2016}. In this work, we present a variety of kirigami actuators whose dynamical pattern formation is controllable. We develop a novel form of non-linear control-response relationships in kirigami geometries through the conversion of linear displacement imposed on the boundary of the thin sheet into a range of predictable motions.

The four fundamental modes depicted in figure~\ref{fig1}, namely roll (rotation about $x$--axis), pitch (rotation about $y$--axis), yaw (rotation about $z$--axis), and lift ($z$--axis out--of--plane displacement), arise from linear actuation, and they may in principle be combined to generate any motion in 3D space. To demonstrate this designing goal, we create three orthogonal rotations and a vertical out--of--plane displacement and show the mechanism for understanding how these emerge from the coupled behavior of individual cuts. We provide a theory that captures the main large scale features in the mechanics of these structures, and demonstrate that similar actuators can be realized in suspended 2D materials, such as graphene and MoS$_{2}$~\cite{Qi2014,Blees2015}. Moreover, a full characterization of the out--of--plane displacement that occurs as a result of a single cut in a thin sheet allows us to derive a scaling law that shows a robust link between simulation and experiment on length--scales ranging over six orders of magnitude. Because kirigami actuators are scale--invariant, our findings can be applied to tailor the microstructure and functionality of mechanical metamaterials across the technological spectrum of length scales ranging from the nanoscale (NEMS)~\cite{Qi2014,Blees2015,Cai2016,Han2017}, the microscale (MEMS)~\cite{Song2015, Xu2016, Rogers2016, Baldwin2017}, and the macroscale~\cite{Saito2011,Sareh2012, Zhang2014,Lamoureux2015}. 

\begin{figure}
\begin{center}
\includegraphics[width=1.0\columnwidth]{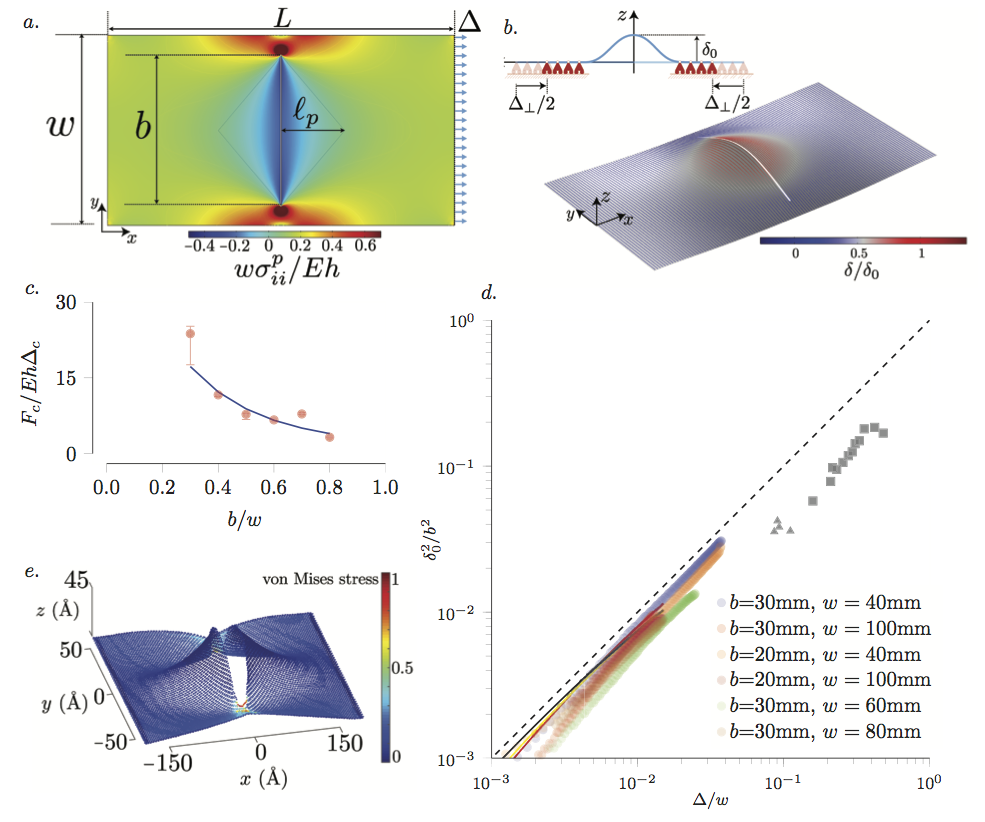}
\end{center}
\vspace{-5mm}
\caption[]{\footnotesize $a.$ Single cut's in--plane state of stress calculated from Finite Element method (FEM). Simulation parameters are set based on the experiments: $h=0.127$mm, $b=80$mm, $w=100$mm, $L=182$mm, Young's Modulus $E = 3.5$GPa, Poisson's ratio $\nu = 0.38$, and $\Delta\sim h$. The color map shows the normalized sum of the principal stresses. $b.$ First mode of deformation, where color map represents the normalized deflection, $\delta/\delta_{\mbox{\tiny 0}}$. $c.$ Critical force $F_c$ required for buckling near the crack as a function of $b/w$. FEM simulation (solid lines) and experimental (disks with error bars) are shown. $d.$ Plot of $\delta_0^2/b^2$ as a function of $\Delta/w$ for experiments with mylar films (circles), FEM simulations (solid lines), and MD simulations of graphene (squares) and MoS2 (triangles). The scaling from equation~\eqref{eq-scaling} is represented by the dashed line. Mylar and FEM parameters are set to $h=0.127$mm, $L=182$mm, $E = 3.5$GPa, and $\nu = 0.38$. MD simulations were done for a fixed $L=346$\AA, we plot date for width $w=114$\AA~and cuts lengths ranging from $b=38$\AA~to 76\AA, and for $w=142$\AA~with cuts ranging from $b=81$\AA~to 119\AA. $e.$ Shows a suspended graphene sheet ($b=76$\AA, $w=114$\AA, $L=346$\AA, $\Delta=40$\AA), where the color map shows the von Mises stress scaled by its maximum value.
\vspace{-3mm}
\label{single-cut}}
\end{figure}

The complex behavior of kirigami actuators arises from functionalizing cracks in thin plates. In other words, when a material is thin enough, cracks under tension may cause the system to buckle before failure through crack propagation~\cite{Hui1998,Brighenti2005}. Therefore, a deeper understanding of the mechanics of a single non--propagating crack on thin sheets is needed. Let us consider a cut of length $b$ centered with respect to the sheet's length $L$ and width $w$, and parallel to the clamped edges of the sheet (figure~\ref{single-cut}$a$). The sheet thickness $h$ is small, such that $h\ll L\sim w$. Applying a uniaxial extension $\Delta$ perpendicular to the crack causes the sheet to buckle out--of--plane at a critical force $F_c$. The typical deflection size is given by a maximum amplitude $\delta_{0}$ centered between the crack tips, and this shape decays back to nearly flat before reaching the clamped boundaries (figure~\ref{single-cut}$b$). This characteristic shape occurs on each side of the crack, such that the shape may be symmetric or antisymmetric about the plane aligned with the crack, normal to the initially flat surface---these two modes, respectively, correspond to stress intensity factors of bending and transverse shear~\cite{Hui1998,Zehnder2005}. We shall here focus our analysis on the symmetric kind, as the typical size of both out--of--plane deformations must be of the same order of magnitude. The critical force needed to trigger this instability is given by $F_c$, which depends on the ratio of the crack to sheet width, $b/w$ (figure~\ref{single-cut}$c$). Since the instability results from an in--plane compressive zone (figure~\ref{single-cut}$a$) around the internal boundary along the crack~\cite{Adda2001,Brighenti2005}, this problem will be approximated by a beam of length $b$. Therefore, $F_c$ is shown to collapse on a single curve (figure~\ref{single-cut}$c$) when the experimental data and simulation results are normalized by the characteristic buckling force $E\,h\,\Delta_c$, where $E$ is the Young's modulus of the material. This will become evident in equation~\eqref{eq-scaling}, where we derive $\Delta_c\equiv h^2/b$ as the critical amount of in--plane compression at the buckling threshold.

\begin{figure}[t]
\begin{center}
\includegraphics[width=1.0\columnwidth]{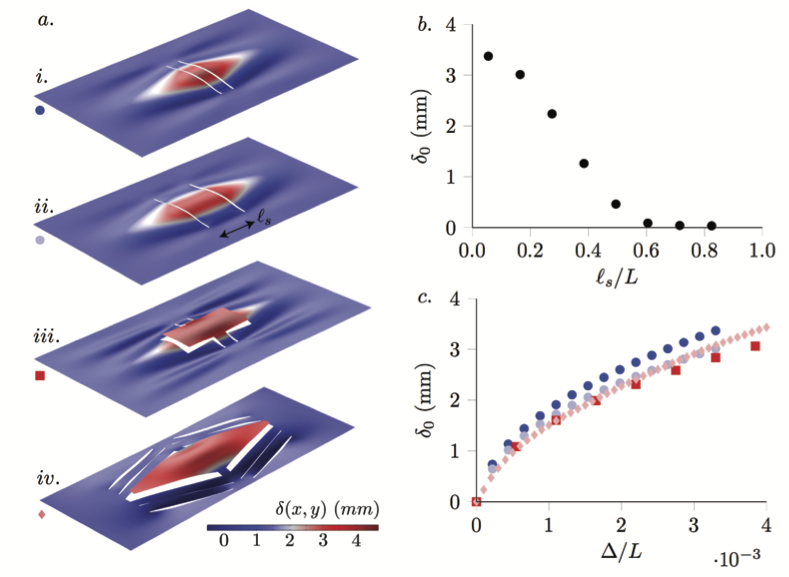}
\end{center}
\vspace{-5mm}
\caption[]{ \footnotesize  $a.$ FEM simulations using two cuts as the basis for generating lift. $b.$ Increasing the spacing between two cuts causes the lift of the center of the sheet to significantly drop. $c.$ By using additional cuts in $iii.$ and $iv.$, we can generate the same lift as with $ii.$ while lifting a much larger area.\label{fig_lift}}
\end{figure}

To describe the post--buckled shape, we consider two regimes: the in--plane stretching dominated response to an applied extension $\Delta$ normal to the single cut, and the out--of--plane state, where the buckling threshold is reached in the stress relief zone and the system becomes bending dominated. This loading condition induces a crack opening mode described by the mode I stress intensity factor, $K_I$, which for a large plate in a state of uniform uniaxial stress is $K_I=T\sqrt{\pi b/2}$, where $T$ is the tensile stress acting on the edge of the sheet~\cite{Bazant2010}. From fracture mechanics~\cite{Bazant2010}, it is stablished that the stress scales with the radius of curvature $r$ of the cut: $\sigma=K_I/\sqrt{2\pi r}$. To estimate the tension in the sheet, we note that stresses concentrate near the crack and, in view of St. Venant's Principle~\cite{Muskhelishvili2013}, it approaches an average value at a distance of about width $w$ away from the crack. This approximation is also validated from the fact that beyond a sheet length to width ratio of about $L/w\approx1$, the maximum deflection of the crack reaches a constant value indicating that, beyond a certain point, the sheet length does not contribute to the crack deformation. Therefore, we expect the tension in the sheet to scale as $T\sim E \Delta/w$. In our experiments we take $r \sim h$ for the crack radius. Therefore, the stress in the sheet becomes $\sigma\sim E (\Delta/w) (b/h)^{1/2}.$ The elastic strain energy due to stretching scales as $\mathcal{U}_s\sim h\left(\sigma^2/E\right) A_s$, where $A_s=L w$ is the area of the sheet, which reduces to
\begin{equation}
\label{eq-stretch}
\mathcal{U}_s \sim E \frac{\Delta^2}{w}b L.
\end{equation}
If we consider the sheet to be dominated by stretching, {\em i.e.} by initially neglecting bending energy, the total potential energy is given as $\mathcal{V}=\mathcal{U}_s-\mathcal{W}$, where $\mathcal{W}$ is the work done by the extension $\Delta$. Taking the work as the force $\left(E\gamma^{\mbox{\tiny$(0)$}}\right)A_s$ times the extension $\Delta$, where $\gamma^{\mbox{\tiny$(0)$}}$ is lateral strain of the sheet, and minimizing the total potential energy, $\left(\partial/\partial\Delta\right)\left[E (\Delta^2/w)b L-E\gamma^{\mbox{\tiny$(0)$}} Lw\Delta\right]=0$, gives a relation for the lateral contraction, 
\begin{equation}
\label{eq-contraction}
\Delta_\perp\equiv\gamma^{\mbox{\tiny$(0)$}} w\sim b\Delta/w.
\end{equation}

Note that equation~\eqref{eq-contraction} is effectively a scaling of Poisson's contraction and sets up the base state for the in--plane solution. We now calculate the next order contribution by allowing the stresses in the compressive zone to reduce the total energy through out--of--plane bending. The calculation is simplified by treating the problem as a 1D buckling of the free boundary along the crack (figure~\ref{single-cut}$b$), where both stretching and bending energies are required to provide the right balance. This next order contribution is obtained as a minimizer of a dimensionally reduced model, along the arc-length $s$ of the cut, given by
\begin{equation}
\label{eq-S-plus-B}
\mathcal{U}=\frac{b\,h\,E}{2}\int\mathrm{d}s\left[\gamma^2+h^2\delta^{\prime\prime\,2}\right],
\end{equation}
where the new measure of strain is geometrically non-linear, $\gamma\approx\gamma^{\mbox{\tiny$(0)$}}+\delta^{\prime\,2}/2$, and $\delta$ is the deflection. This yields a classic result for the maximum amplitude:
\begin{equation}
\label{eq-deflection}
\delta_{0}\sim\sqrt{b}\,\sqrt{\Delta_{\mbox{\tiny$\perp$}}-\Delta_c},
\end{equation}
where $\Delta_c$ is related to the ratio between bending rigidity, $B=h^3E$, and stretching rigidity, $Y=hE$, as follows: $\Delta_c/b\sim B/\left(b^2\,Y\right)=\left(h/b\right)^2$.
Inserting the in--plane compression result of equation~\eqref{eq-contraction} into \eqref{eq-deflection} gives a scaling for the maximum crack deflection,
\begin{equation}
\label{eq-scaling}
\left(\frac{\delta_{0}}{b}\right)^2\sim\frac{\Delta}{w}-\mathcal{O}\left(\frac{h}{b}\right)^2.
\end{equation}
Equation~\ref{eq-scaling} shows a higher order dependency on the sheet thickness to crack length ratio, implying the invariance of these deformations from the macro to the nanoscale. To confirm this relationship, experiments were performed with single cuts in mylar films (Biaxially-oriented polyethylene terephthalate---BoPET) to measure the maximum deflection as a function of extension for a given crack size and sheet width (see methods section). Finite Element Method (FEM) simulations with the same material parameters were also performed (see methods section). Additionally, we carried out Molecular Dynamics (MD) simulations of suspended graphene monolayers (see methods section). Figure~\ref{single-cut}$d$ shows the dimensionless deflection data for the experiments and simulations, along with the scaling prediction from equation \ref{eq-scaling}, confirming a very strong agreement across six orders of magnitude.

In order to generate simple actuators that can become the building blocks for more complex structures, such as mechanical metamaterials, we must quantify how multiple cracks will interact to generate motion of points on the sheet. Since the behavior of a single crack is well described by equation~\eqref{eq-scaling}, the simplest extension is two parallel cracks of length $b$ separated by distance $\ell_s$. When $\ell_s/L$ is small, these cracks interact to generate vertical lift of the sheet between them (figure~\ref{fig_lift}$a$--$i$ \& $ii$). However, the deflection of the center point of the sheet drops off quickly as the spacing between the cracks is increased, making it difficult to lift a large amount of surface area (figure~\ref{fig_lift}$b$). Keeping $\ell_s/L$ small while increasing the area of the sheet that is lifted can be accomplished by extending a portion of each crack towards the clamped boundaries (figure~\ref{fig_lift}$a$--$iii$). This relies on the same buckling mechanism that governs the single crack behavior, producing nearly the same amount of lift as the two parallel cracks (figure~\ref{fig_lift}$c$). These additional cuts also introduce wrinkles on the sheet, which can be avoided by introducing cuts that provide room for in--plane compression (figure~\ref{fig_lift}$a$--$iv$). With this arrangement of cuts, we demonstrate how these parallel cracks can become building blocks for generating lift of a large, localized area.

\begin{figure}[t]
\begin{center}
\includegraphics[width=1.0\columnwidth]{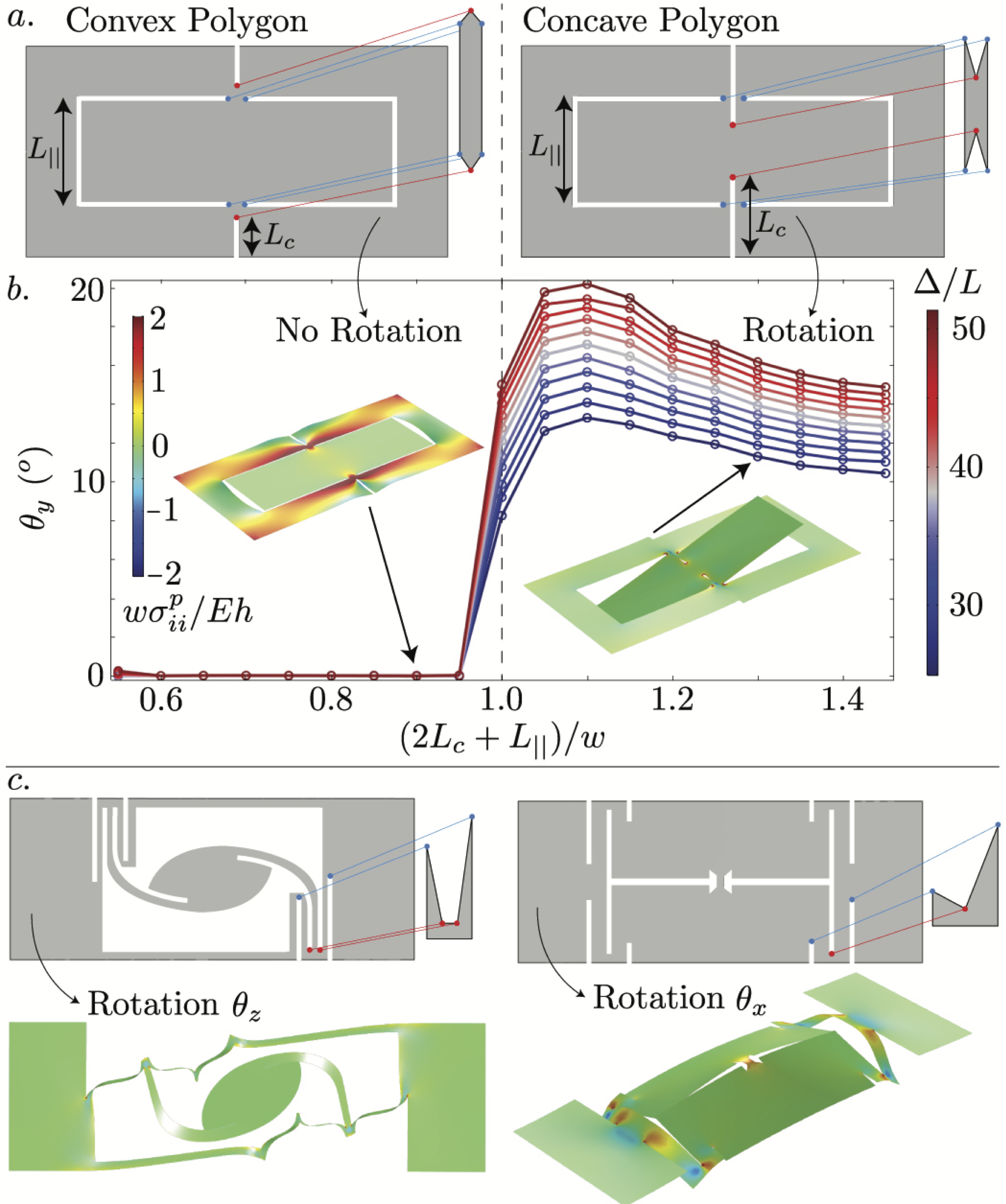}
\end{center}
\vspace{-5mm}
\caption[]{\footnotesize  $a.$ Schematics illustrating how the polygon formed by crack tips will generate rotation. $b.$ As the polygon formed by the edge crack tip and the internal crack tips changes from convex to concave, we see the emergence of rotation about the $y$--axis. $c.$ The coupling of multiple concave polygons formed by the crack tips can enable rotation about the $x$ or $z$ axis as well.\vspace{-3mm}
\label{fig_rotate}}
\end{figure}

We note that the four crack tips of the two parallel cracks in figure~\ref{fig_lift}$a$--$i$ form a rectangular {\em unit cell} (convex polygon) and generate lift in the sheet. This rudimentary shape is identified quantitatively by following the lines of tension that connect two neighboring cracks, and the convexity of the unit cell signifies how much stretching within the sheet can be transferred into a crack opening displacement. Convex shapes constrain the sheet to induce vertical lift, while concave shapes have the freedom to rotate. To illustrate this idea, we focus on the pitch mode. We performed a post--buckling analysis through FEM simulations for the geometry in figure~\ref{fig_rotate}$a$, while varying crack length $L_c$, thus allowing us to scan unit cell shapes from convex to concave. Denoting $L_{\parallel}$ as the cut length parallel to the clamped boundary, we refer to the ratio $(2L_c+L_{\parallel})/w$ as a measure of convexity. The target shape strongly depends on this parameter's transition: lift of the outer portion of the sheet occurs when the unit cell is convex, {\em i.e.} $(2L_c+L_{\parallel})/w<1$, while rotation about the $y$--axis occurs when it is concave, {\em i.e.} $(2L_c+L_{\parallel})/w\gtrsim1$ (figure~\ref{fig_rotate}$b$). Generating rotation about the $z$ and $x$ axes follows the same principle---concave unit cells enable rotation (figure~\ref{fig_rotate}$c$). In these more complex configurations, there is coupling between two unit cells within the sheet. While an intricate model of the coupling between multiple unit cells is beyond the scope of this work, it is clear from the schematics and post--buckled shapes that the concave unit cells locally enable rotation about the $z$ and $x$ axes.

\begin{figure}
\begin{center}
\includegraphics[width=1.0\columnwidth]{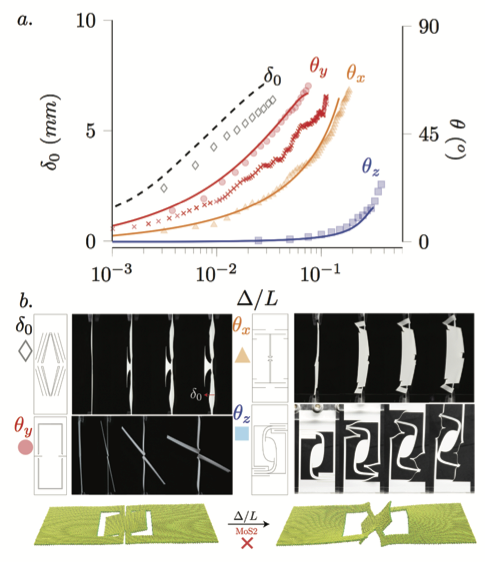}
\end{center}
\vspace{-5mm}
\caption[]{\footnotesize $a.$ A plot of the lift of the center $\delta_0$ (black) and the three rotations as functions of $\Delta/L$, $\Delta$ being the applied displacement in the $x$--direction. The experimental data corresponds to black diamond for lift, orange triangles for the roll (rotation about $x$--axis), red disks for the pitch (rotation about $y$--axis), and blue squares for the yaw (rotation about $z$--axis). FEM for the respective modes of deformation, using the same parameters of the experiments, are shown in dashed and solid curves. The red x's show the results of the molecular dynamics simulations. $b.$ Images of the experiments for the cut patterns and the sequence of deformation as $\Delta/L$ is increased as well as two snapshots of the molecular dynamics simulation. In the case of the MoS$_{2}$, the geometric parameters are: length $L=$460\AA and width $w=$152\AA of the sheet; crack length $L_c=47.5$\AA; and a 240\AA~length and a 82\AA~width ($L_{\parallel}$) of the inner rotating ribbon.\label{fig-lift-rot}}
\end{figure}

Figure~\ref{fig_rotate} indicates that the convexity of the unit cell formed by the locally interacting crack tips can generate either lift or rotation. We provide further insight through quantifying the magnitude of these kirigami--based motions by measuring the lift or rotation as a function of relative strain $\Delta/L$ (figure~\ref{fig-lift-rot}). Here we show that a portion of the sheet can achieve a vertical displacement nearly 50 times the sheet thickness. Since there is no plastic deformation and the cracks do not propagate, these deformations are reversible. The stiffness of the sheets designed to provide rotation varies widely. Rotations about the $y$ (pitch) and $x$ (roll) axes reach about 60 degrees after a moderate amount of extension, while the in--plane rotation about $z$--axis requires a significant amount of extension to reach 30 degrees of rotation. Figure~\ref{fig-lift-rot}$a$ shows good agreement between the experimental measurements for the macroscale designs of lift (diamonds for  $\delta_0$) and rotation (triangles for $\theta_x$, disks for $\theta_y$, and squares for $\theta_z$) and the FEM simulations (dashed line for $\delta_0$, orange for $\theta_x$, red for $\theta_y$, and blue for $\theta_z$). 

The results from figure~\ref{single-cut}$d$ suggest that these actuator designs should scale down to 2D materials. From a MoS$_{2}$ monolayer, we tested the simplest nanoactuator requiring only one unit cell for rotation about the $y$--axis, \emph{i.e.} the pitch mode shown at the bottom in figure~\ref{fig-lift-rot}$b$. We obtained a rigid rotation due to its higher bending modulus than that of graphene~\cite{jiang-Nanotechnology-24-435705-2013}. We applied an extension perpendicular to the crack and measured the rotation of the inner ribbon about the $y$--axis (figure~\ref{fig-lift-rot}$a$, red x's). At small $\Delta/L$, there is good agreement between the macroscale results and the nanoscale simulations, and eventually the three actuators achieve nearly the same maximum value of $\theta_y$. While the behavior is qualitatively similar across several orders of magnitude in sheet thickness, it is clear that the agreement for the 2D kirigami is qualitative rather than quantitative. Specifically, the fact that the rotation that is observed in the 2D kirigami is smaller for the same strains than the bulk system suggests that the 2D system may undergo more stretching than the bulk system, a point also made recently by Grosso and Mele~\cite{Grosso2015}. Therefore, additional analysis of the 2D material kirigami actuators is necessary to quantitatively replicate the macroscale actuator designs.

Finally, we return to the actuators in figure~\ref{fig1}. Through replicating the mechanism in figure~\ref{fig_rotate}$a$, we see rotation about the $y$--axis of all cells (Figure~\ref{fig1}$a$). This indicates that building blocks can go beyond mechanism design towards the development of mechanical metamaterials. Furthermore, the interactions between multiple cuts can enable portions of a thin sheet to rotate one complete revolution about the $x$--axis the extension axis (figure~\ref{fig1}$c$), while coupling unit cells that cause rotation and lift generates sheets that first rotate about the $y$--axis and subsequently lift in the $z$ direction (figure~\ref{fig1}$d$). What remains is to better understand how building blocks can be combined to generate targeted behaviors---an inverse problem that can begin by considering the simple geometric model we present here.

We have addressed two fundamental problems that are pivotal to connecting kirigami actuators to practical designs for engineering applications: scale-invariant behavior and a robust geometric mechanism for actuator design. While the kirigami mechanics has been unified over six orders of magnitude in sheet thickness, the shape of a unit cell formed by locally interacting crack tips provides a geometric mechanism to induce either lift or rotation. What we present has the potential to offer rational designing tools for dynamical assembling of complex geometries~\cite{Zhang2015,Lamoureux2015}, and we hope that this spontaneous generation of shapes emerging from quasi-static actuation comes to complement inverse design algorithms that have been proposed for lattice-based kirigami~\cite{Castle2014,Sussman2015}. As it has been previously mentioned, the cracks do not propagate in the experiments performed here, thus the process remains entirely reversible. In order to maintain this reversibility in systems utilizing materials with lower yield stress, cracks can be made with a larger crack tip radius $r$, thus lowering the stress intensity factor, $K_I$. The scaling found here is robust under such a modification since a few multiples of $r$ only yield a pre-factor in front of equation~\eqref{eq-scaling}, thus preserving the same power-law. It is also noted that the propagation of interacting cracks can be manipulated by their initial geometry~\cite{Cortet2008,Dalbe2015}, such interactions could be utilized to increase the functionality of the kirigrami structures and/or give a predictable response to strain beyond that which causes crack propagation. There may also be significant scientific benefits to demonstrating kirigami actuation in 2D materials. From a basic science perspective, kirigami provides an ideal platform to study the localization of electronic states, or the coupling of 2D quantum dots~\cite{Bahamon2016}.  Alternatively, these structures offer significant opportunities for flexible, lightweight band-gap engineered optoelectronic materials whose performance can be reversibly changed and manipulated over a wide range of the optical spectrum by locally varying the strain~\cite{Castellanos2013,Tsai2015,He2013,Lloyd2016,Feng2012,Conley2013,Johari2012,Kotov2012}.

\section*{Methods Section}

{\em Fabrication:} Mylar (BoPET) films were purchased from McMaster--Carr (Mylar, 8567K96), and had a thickness of $h=0.127$ mm.
To relieve any residual stress in the films, apparent from their natural curvature, we annealed the films in the oven at 85$^{o}$C under the weight of thick metal sheets for 2 hours, resulting in flat sheets. Vector patterns were drawn in Adobe Illustrator CS6, and cut with an Epilog Mini 24, 75W laser cutter in vector mode, at 80$\%$ speed and 10$\%$ power. Sheet widths of $w=40$mm, 60mm, 80mm, and 100mm were used, and sheet lengths of $L=20$mm to 200mm in linear increments of 20mm were used. For the single cut experiments, cut lengths ranging from $b=20$mm to 70mm were used. The cut mylar films were adhered to 3mm thick acrylic sheets (McMaster--Carr, acrylic, 8560K191) with cyanoacrylate glue (McMaster-Carr, Loctite 403, 74765A53), which served as the clamped boundary conditions for the films.  

{\em Mechanical Measurements:} Uniaxial tension tests were performed by clamping the mylar sheets to the Instron 5943 mechanical testing system, using a 500N load cell. Displacement--controlled tests were performed at a rate of 0.15mm/min to a maximum extension of 1.5mm. Since the mylar did not experience inelastic strains, actuation was reversible, and 3 tests were run for each sample. Actuator deformation was captured from the side  with a microscopic lens (Navitar Zoom 6000) attached to a Nikon D610 camera, and from the front using a Nikon D610 camera with a Micro-NIKKOR 105mm f/2.8 Lens, a Nikon 55mm f/2.8 Lens, and a high contrast Rosco Color Filter (B\&H Photo Video, ROCEK1212). The critical buckling force was determined from identifying both the slope change in the force vs. displacement curve, and the out--of--plane deflection from the microscopic imaging of the crack profile.

{\em Finite Element Method (FEM):} FEM simulations were undertaken using COMSOL Multiphysics 5.2~\cite{COMSOL} along with the Structural Mechanics Module. Shell Mechanics and Plates were the environments within COMSOL in which all of our studies were performed. A geometry matching those used in the experiment was created in COMSOL's Design Module. Mesh refinement studies were undertaken to ensure convergence of the results. For the single cut geometry in figure~2, the sheet was modeled as an isotropic elastic thin sheet with thickness of $h = 0.127$mm, Young's Modulus $E = 3.5$GPa, and Poisson's ratio $\nu = 0.38$. The results shown in figure~2$c$ were attained through linear buckling studies with varying thickness in the range $h\in\left[0.1\mathrm{mm},0.14\mathrm{mm}\right]$ and the values $b/w\in\left[0.3,0.8\right]$. The in--plane results shown in figure~2$a$ and post-buckling results in figures~2$d$, 3, and 4 were calculated from a stationary study with displacement ($\Delta$) controlled analysis. In order to induce out--of--plane symmetry breaking, we added random small imperfections (ten orders of magnitude smaller than the sheet thickness) to the initial surface. The parameters in figure~2$d$ varied and lay in the ranges $h\in\left[0.15\mathrm{mm},0.21\mathrm{mm}\right]$ and $b/w\in\left[0.1,0.9\right]$. For the results shown in figure~5, linear buckling studies were undertaken with the boundary at $x=0$ fixed in space while the boundary at $x=L$ had an imposed displacement of $\Delta$ in the $x$ direction. Both of these boundaries were not permitted to rotate. All other boundaries were free. Small imperfections in the form of the first eigenmodes were then added to the initially flat geometry through use of MeshPerturb 1.0 \cite{saha2014meshperturb}. These imperfect geometries were then used for the stationary studies with the same boundary conditions and the same mesh density as was used in the linear studies. 

{\em Molecular Simulations:} We used Sandia-developed open source LAMMPS Molecular Dynamics Simulator to simulate graphene sheets~\cite{plimptonLAMMPS}. To describe the carbon-carbon interactions, we used AIREBO potential~\cite{stuart2000reactive} as has been used previously in atomistic study of graphene kirigami~\cite{Qi2014}. The cutoffs for the Lennard-Jones and the REBO term in AIREBO potential are chosen to be 2~\AA~and 6.8~\AA, respectively. For MoS$_{2}$ actuators we used the Stillinger-Weber potential developed by Jiang~\cite{jiang-Nanotechnology-26-315706-2015}, which we have previously employed to study MoS$_{2}$ kirigami~\cite{hanakata-Nanoscale-8-458-2016}. Graphene with a single crack and the MoS$_{2}$ actuator were first relaxed for 50--200~ps at 4.2K within the NVT (fixed number of atoms $N$, volume $V$, and temperature $T$) ensemble. Non-periodic boundary conditions were applied in all three directions. After the relaxation, the strains were applied by displacing  both ends at a uniform rate. 

\section*{Author contributions} 
M.A.D. and D.P.H. conceived the study and proposed the research; M.A.D. and D.R.K. performed the
mathematical modeling and FEM in consultation with D.P.H.; M.P.M. and D.P.H. performed the experiments in consultation with M.A.D.; P.Z.H., D.K.C., and H.S.P. performed the Molecular Simulations studies; M.A.D. and D.P.H. wrote the main article; and all authors jointly edited the entire article.

\section*{Acknowledgements} 
M.A.D. is grateful to M. Adda-Bedia, J. Bico, B. Roman, and K. Seffen for many insightful discussions. M.A.D. would also like to thank ENS-Lyon and ESPCI-Paris for hosting the author during part of the development of this manuscript and funding from the 4-VA program for collaborative research at JMU. D.P.H. is grateful to the National Science Foundation (CMMI CAREER--1454153) for financial support. M.A.D., D.R.K., and D.P.H. would like to thank I. Metta for the useful discussions at the start of this project. D.K.C. acknowledges the Aspen Center for Physics, which is supported by National Science Foundation grant PHY-1607611, where part of this work was completed. D.R.K. acknowledges funding from Academy of Finland.

\section*{Conflict of interest}
The authors declare no conflict of interest.






\begin{thebibliography}{53}
\expandafter\ifx\csname natexlab\endcsname\relax\def\natexlab#1{#1}\fi
\expandafter\ifx\csname bibnamefont\endcsname\relax
  \def\bibnamefont#1{#1}\fi
\expandafter\ifx\csname bibfnamefont\endcsname\relax
  \def\bibfnamefont#1{#1}\fi
\expandafter\ifx\csname citenamefont\endcsname\relax
  \def\citenamefont#1{#1}\fi
\expandafter\ifx\csname url\endcsname\relax
  \def\url#1{\texttt{#1}}\fi
\expandafter\ifx\csname urlprefix\endcsname\relax\def\urlprefix{URL }\fi
\providecommand{\bibinfo}[2]{#2}
\providecommand{\eprint}[2][]{\url{#2}}

\bibitem[{\citenamefont{Reis}(2015)}]{Reis2015}
\bibinfo{author}{\bibfnamefont{P.~M.} \bibnamefont{Reis}},
  \bibinfo{journal}{Journal of Applied Mechanics}
  \textbf{\bibinfo{volume}{82}}, \bibinfo{pages}{111001}
  (\bibinfo{year}{2015}).

\bibitem[{\citenamefont{Seffen}(2016)}]{Seffen2016}
\bibinfo{author}{\bibfnamefont{K.~A.} \bibnamefont{Seffen}},
  \bibinfo{journal}{Physical Review E} \textbf{\bibinfo{volume}{94}},
  \bibinfo{pages}{033003} (\bibinfo{year}{2016}).

\bibitem[{\citenamefont{Rafsanjani and Bertoldi}(2017)}]{Rafsanjani2017}
\bibinfo{author}{\bibfnamefont{A.}~\bibnamefont{Rafsanjani}} \bibnamefont{and}
  \bibinfo{author}{\bibfnamefont{K.}~\bibnamefont{Bertoldi}},
  \bibinfo{journal}{Physical Review Letters} \textbf{\bibinfo{volume}{118}},
  \bibinfo{pages}{084301} (\bibinfo{year}{2017}).

\bibitem[{\citenamefont{Bico et~al.}(2017)\citenamefont{Bico, Lepoivre, Bense,
  Reyssat, and Roman}}]{Bico2017}
\bibinfo{author}{\bibfnamefont{J.}~\bibnamefont{Bico}},
  \bibinfo{author}{\bibfnamefont{E.}~\bibnamefont{Lepoivre}},
  \bibinfo{author}{\bibfnamefont{H.}~\bibnamefont{Bense}},
  \bibinfo{author}{\bibfnamefont{E.}~\bibnamefont{Reyssat}}, \bibnamefont{and}
  \bibinfo{author}{\bibfnamefont{B.}~\bibnamefont{Roman}},
  \bibinfo{journal}{Bulletin of the American Physical Society}
  \textbf{\bibinfo{volume}{62}} (\bibinfo{year}{2017}).

\bibitem[{\citenamefont{Virk et~al.}(2013)\citenamefont{Virk, Monti, Trehard,
  Marsh, Hazra, Boba, Remillat, Scarpa, and Farrow}}]{Virk2013}
\bibinfo{author}{\bibfnamefont{K.}~\bibnamefont{Virk}},
  \bibinfo{author}{\bibfnamefont{A.}~\bibnamefont{Monti}},
  \bibinfo{author}{\bibfnamefont{T.}~\bibnamefont{Trehard}},
  \bibinfo{author}{\bibfnamefont{M.}~\bibnamefont{Marsh}},
  \bibinfo{author}{\bibfnamefont{K.}~\bibnamefont{Hazra}},
  \bibinfo{author}{\bibfnamefont{K.}~\bibnamefont{Boba}},
  \bibinfo{author}{\bibfnamefont{C.~D.~L.} \bibnamefont{Remillat}},
  \bibinfo{author}{\bibfnamefont{F.}~\bibnamefont{Scarpa}}, \bibnamefont{and}
  \bibinfo{author}{\bibfnamefont{I.~R.} \bibnamefont{Farrow}},
  \bibinfo{journal}{Smart Materials and Structures}
  \textbf{\bibinfo{volume}{22}}, \bibinfo{pages}{084014}
  (\bibinfo{year}{2013}).

\bibitem[{\citenamefont{Scarpa et~al.}(2013)\citenamefont{Scarpa, Ouisse,
  Collet, and Saito}}]{Scarpa2013}
\bibinfo{author}{\bibfnamefont{F.}~\bibnamefont{Scarpa}},
  \bibinfo{author}{\bibfnamefont{M.}~\bibnamefont{Ouisse}},
  \bibinfo{author}{\bibfnamefont{M.}~\bibnamefont{Collet}}, \bibnamefont{and}
  \bibinfo{author}{\bibfnamefont{K.}~\bibnamefont{Saito}},
  \bibinfo{journal}{Journal of Vibration and Acoustics}
  \textbf{\bibinfo{volume}{135}}, \bibinfo{pages}{041001}
  (\bibinfo{year}{2013}).

\bibitem[{\citenamefont{Tang and Yin}(2017)}]{Tang2017}
\bibinfo{author}{\bibfnamefont{Y.}~\bibnamefont{Tang}} \bibnamefont{and}
  \bibinfo{author}{\bibfnamefont{J.}~\bibnamefont{Yin}},
  \bibinfo{journal}{Extreme Mechanics Letters} \textbf{\bibinfo{volume}{12}},
  \bibinfo{pages}{77} (\bibinfo{year}{2017}).

\bibitem[{\citenamefont{Shyu et~al.}(2015)\citenamefont{Shyu, Damasceno, Dodd,
  Lamoureux, Xu, Shlian, Shtein, Glotzer, and Kotov}}]{Shyu2015}
\bibinfo{author}{\bibfnamefont{T.~C.} \bibnamefont{Shyu}},
  \bibinfo{author}{\bibfnamefont{P.~F.} \bibnamefont{Damasceno}},
  \bibinfo{author}{\bibfnamefont{P.~M.} \bibnamefont{Dodd}},
  \bibinfo{author}{\bibfnamefont{A.}~\bibnamefont{Lamoureux}},
  \bibinfo{author}{\bibfnamefont{L.}~\bibnamefont{Xu}},
  \bibinfo{author}{\bibfnamefont{M.}~\bibnamefont{Shlian}},
  \bibinfo{author}{\bibfnamefont{M.}~\bibnamefont{Shtein}},
  \bibinfo{author}{\bibfnamefont{S.~C.} \bibnamefont{Glotzer}},
  \bibnamefont{and} \bibinfo{author}{\bibfnamefont{N.~A.} \bibnamefont{Kotov}},
  \bibinfo{journal}{Nature materials} \textbf{\bibinfo{volume}{14}},
  \bibinfo{pages}{785} (\bibinfo{year}{2015}).

\bibitem[{\citenamefont{Vachicouras et~al.}(2017)\citenamefont{Vachicouras,
  Tringides, Campiche, and Lacour}}]{Vachicouras2017}
\bibinfo{author}{\bibfnamefont{N.}~\bibnamefont{Vachicouras}},
  \bibinfo{author}{\bibfnamefont{C.~M.} \bibnamefont{Tringides}},
  \bibinfo{author}{\bibfnamefont{P.~B.} \bibnamefont{Campiche}},
  \bibnamefont{and} \bibinfo{author}{\bibfnamefont{S.~P.}
  \bibnamefont{Lacour}}, \bibinfo{journal}{Extreme Mechanics Letters}
  \textbf{\bibinfo{volume}{15}}, \bibinfo{pages}{63 } (\bibinfo{year}{2017}).

\bibitem[{\citenamefont{Zhang et~al.}(2015)\citenamefont{Zhang, Yan, Nan, Xiao,
  Liu, Luan, Fu, Wang, Yang, Wang et~al.}}]{Zhang2015}
\bibinfo{author}{\bibfnamefont{Y.}~\bibnamefont{Zhang}},
  \bibinfo{author}{\bibfnamefont{Z.}~\bibnamefont{Yan}},
  \bibinfo{author}{\bibfnamefont{K.}~\bibnamefont{Nan}},
  \bibinfo{author}{\bibfnamefont{D.}~\bibnamefont{Xiao}},
  \bibinfo{author}{\bibfnamefont{Y.}~\bibnamefont{Liu}},
  \bibinfo{author}{\bibfnamefont{H.}~\bibnamefont{Luan}},
  \bibinfo{author}{\bibfnamefont{H.}~\bibnamefont{Fu}},
  \bibinfo{author}{\bibfnamefont{X.}~\bibnamefont{Wang}},
  \bibinfo{author}{\bibfnamefont{Q.}~\bibnamefont{Yang}},
  \bibinfo{author}{\bibfnamefont{J.}~\bibnamefont{Wang}}, \bibnamefont{et~al.},
  \bibinfo{journal}{Proceedings of the National Academy of Sciences}
  \textbf{\bibinfo{volume}{112}}, \bibinfo{pages}{11757}
  (\bibinfo{year}{2015}).

\bibitem[{\citenamefont{Sadoc et~al.}(2012)\citenamefont{Sadoc, Rivier, and
  Charvolin}}]{Sadoc2012}
\bibinfo{author}{\bibfnamefont{J.-F.} \bibnamefont{Sadoc}},
  \bibinfo{author}{\bibfnamefont{N.}~\bibnamefont{Rivier}}, \bibnamefont{and}
  \bibinfo{author}{\bibfnamefont{J.}~\bibnamefont{Charvolin}},
  \bibinfo{journal}{Acta Crystallographica Section A: Foundations of
  Crystallography} \textbf{\bibinfo{volume}{68}}, \bibinfo{pages}{470}
  (\bibinfo{year}{2012}).

\bibitem[{\citenamefont{Charvolin and Sadoc}(2011)}]{Charvolin2011}
\bibinfo{author}{\bibfnamefont{J.}~\bibnamefont{Charvolin}} \bibnamefont{and}
  \bibinfo{author}{\bibfnamefont{J.-F.} \bibnamefont{Sadoc}},
  \bibinfo{journal}{Biophysical Reviews and Letters}
  \textbf{\bibinfo{volume}{6}}, \bibinfo{pages}{13} (\bibinfo{year}{2011}).

\bibitem[{\citenamefont{Sadoc et~al.}(2013)\citenamefont{Sadoc, Charvolin, and
  Rivier}}]{Sadoc2013}
\bibinfo{author}{\bibfnamefont{J.-F.} \bibnamefont{Sadoc}},
  \bibinfo{author}{\bibfnamefont{J.}~\bibnamefont{Charvolin}},
  \bibnamefont{and} \bibinfo{author}{\bibfnamefont{N.}~\bibnamefont{Rivier}},
  \bibinfo{journal}{Journal of Physics A: Mathematical and Theoretical}
  \textbf{\bibinfo{volume}{46}}, \bibinfo{pages}{295202}
  (\bibinfo{year}{2013}).

\bibitem[{\citenamefont{Castle et~al.}(2014)\citenamefont{Castle, Cho, Gong,
  Jung, Sussman, Yang, and Kamien}}]{Castle2014}
\bibinfo{author}{\bibfnamefont{T.}~\bibnamefont{Castle}},
  \bibinfo{author}{\bibfnamefont{Y.}~\bibnamefont{Cho}},
  \bibinfo{author}{\bibfnamefont{X.}~\bibnamefont{Gong}},
  \bibinfo{author}{\bibfnamefont{E.}~\bibnamefont{Jung}},
  \bibinfo{author}{\bibfnamefont{D.~M.} \bibnamefont{Sussman}},
  \bibinfo{author}{\bibfnamefont{S.}~\bibnamefont{Yang}}, \bibnamefont{and}
  \bibinfo{author}{\bibfnamefont{R.~D.} \bibnamefont{Kamien}},
  \bibinfo{journal}{Physical Review Letters} \textbf{\bibinfo{volume}{113}},
  \bibinfo{pages}{1} (\bibinfo{year}{2014}).

\bibitem[{\citenamefont{Sussman et~al.}(2015)\citenamefont{Sussman, Cho,
  Castle, Gong, Jung, Yang, and Kamien}}]{Sussman2015}
\bibinfo{author}{\bibfnamefont{D.~M.} \bibnamefont{Sussman}},
  \bibinfo{author}{\bibfnamefont{Y.}~\bibnamefont{Cho}},
  \bibinfo{author}{\bibfnamefont{T.}~\bibnamefont{Castle}},
  \bibinfo{author}{\bibfnamefont{X.}~\bibnamefont{Gong}},
  \bibinfo{author}{\bibfnamefont{E.}~\bibnamefont{Jung}},
  \bibinfo{author}{\bibfnamefont{S.}~\bibnamefont{Yang}}, \bibnamefont{and}
  \bibinfo{author}{\bibfnamefont{R.~D.} \bibnamefont{Kamien}},
  \bibinfo{journal}{Proceedings of the National Academy of Sciences} p.
  \bibinfo{pages}{201506048} (\bibinfo{year}{2015}).

\bibitem[{\citenamefont{Chen et~al.}(2016)\citenamefont{Chen, Liu, Evans,
  Paulose, Cohen, Vitelli, and Santangelo}}]{Chen2016}
\bibinfo{author}{\bibfnamefont{B.~G.-g.} \bibnamefont{Chen}},
  \bibinfo{author}{\bibfnamefont{B.}~\bibnamefont{Liu}},
  \bibinfo{author}{\bibfnamefont{A.~A.} \bibnamefont{Evans}},
  \bibinfo{author}{\bibfnamefont{J.}~\bibnamefont{Paulose}},
  \bibinfo{author}{\bibfnamefont{I.}~\bibnamefont{Cohen}},
  \bibinfo{author}{\bibfnamefont{V.}~\bibnamefont{Vitelli}}, \bibnamefont{and}
  \bibinfo{author}{\bibfnamefont{C.~D.} \bibnamefont{Santangelo}},
  \bibinfo{journal}{Physical review letters} \textbf{\bibinfo{volume}{116}},
  \bibinfo{pages}{135501} (\bibinfo{year}{2016}).

\bibitem[{\citenamefont{Qi et~al.}(2014)\citenamefont{Qi, Campbell, and
  Park}}]{Qi2014}
\bibinfo{author}{\bibfnamefont{Z.}~\bibnamefont{Qi}},
  \bibinfo{author}{\bibfnamefont{D.~K.} \bibnamefont{Campbell}},
  \bibnamefont{and} \bibinfo{author}{\bibfnamefont{H.~S.} \bibnamefont{Park}},
  \bibinfo{journal}{Physical Review B} \textbf{\bibinfo{volume}{90}},
  \bibinfo{pages}{245437} (\bibinfo{year}{2014}).

\bibitem[{\citenamefont{Blees et~al.}(2015)\citenamefont{Blees, Barnard, Rose,
  Roberts, McGill, Huang, Ruyack, Kevek, Kobrin, Muller et~al.}}]{Blees2015}
\bibinfo{author}{\bibfnamefont{M.~K.} \bibnamefont{Blees}},
  \bibinfo{author}{\bibfnamefont{A.~W.} \bibnamefont{Barnard}},
  \bibinfo{author}{\bibfnamefont{P.~A.} \bibnamefont{Rose}},
  \bibinfo{author}{\bibfnamefont{S.~P.} \bibnamefont{Roberts}},
  \bibinfo{author}{\bibfnamefont{K.~L.} \bibnamefont{McGill}},
  \bibinfo{author}{\bibfnamefont{P.~Y.} \bibnamefont{Huang}},
  \bibinfo{author}{\bibfnamefont{A.~R.} \bibnamefont{Ruyack}},
  \bibinfo{author}{\bibfnamefont{J.~W.} \bibnamefont{Kevek}},
  \bibinfo{author}{\bibfnamefont{B.}~\bibnamefont{Kobrin}},
  \bibinfo{author}{\bibfnamefont{D.~A.} \bibnamefont{Muller}},
  \bibnamefont{et~al.}, \bibinfo{journal}{Nature}
  \textbf{\bibinfo{volume}{524}}, \bibinfo{pages}{204} (\bibinfo{year}{2015}).

\bibitem[{\citenamefont{Cai et~al.}(2016)\citenamefont{Cai, Luo, Ling, Wan, and
  Qin}}]{Cai2016}
\bibinfo{author}{\bibfnamefont{K.}~\bibnamefont{Cai}},
  \bibinfo{author}{\bibfnamefont{J.}~\bibnamefont{Luo}},
  \bibinfo{author}{\bibfnamefont{Y.}~\bibnamefont{Ling}},
  \bibinfo{author}{\bibfnamefont{J.}~\bibnamefont{Wan}}, \bibnamefont{and}
  \bibinfo{author}{\bibfnamefont{Q.-H.} \bibnamefont{Qin}},
  \bibinfo{journal}{Scientific Reports} \textbf{\bibinfo{volume}{6}}
  (\bibinfo{year}{2016}).

\bibitem[{\citenamefont{Han et~al.}(2017)\citenamefont{Han, Scarpa, and
  Allan}}]{Han2017}
\bibinfo{author}{\bibfnamefont{T.}~\bibnamefont{Han}},
  \bibinfo{author}{\bibfnamefont{F.}~\bibnamefont{Scarpa}}, \bibnamefont{and}
  \bibinfo{author}{\bibfnamefont{N.~L.} \bibnamefont{Allan}},
  \bibinfo{journal}{Thin Solid Films}  (\bibinfo{year}{2017}).

\bibitem[{\citenamefont{Song et~al.}(2015)\citenamefont{Song, Wang, Lv, An,
  Liang, Ma, He, Zheng, Huang, Yu et~al.}}]{Song2015}
\bibinfo{author}{\bibfnamefont{Z.}~\bibnamefont{Song}},
  \bibinfo{author}{\bibfnamefont{X.}~\bibnamefont{Wang}},
  \bibinfo{author}{\bibfnamefont{C.}~\bibnamefont{Lv}},
  \bibinfo{author}{\bibfnamefont{Y.}~\bibnamefont{An}},
  \bibinfo{author}{\bibfnamefont{M.}~\bibnamefont{Liang}},
  \bibinfo{author}{\bibfnamefont{T.}~\bibnamefont{Ma}},
  \bibinfo{author}{\bibfnamefont{D.}~\bibnamefont{He}},
  \bibinfo{author}{\bibfnamefont{Y.-J.} \bibnamefont{Zheng}},
  \bibinfo{author}{\bibfnamefont{S.-Q.} \bibnamefont{Huang}},
  \bibinfo{author}{\bibfnamefont{H.}~\bibnamefont{Yu}}, \bibnamefont{et~al.},
  \bibinfo{journal}{Scientific reports} \textbf{\bibinfo{volume}{5}},
  \bibinfo{pages}{10988} (\bibinfo{year}{2015}).

\bibitem[{\citenamefont{Xu et~al.}(2016)\citenamefont{Xu, Wang, Kim, Shyu, Lyu,
  and Kotov}}]{Xu2016}
\bibinfo{author}{\bibfnamefont{L.}~\bibnamefont{Xu}},
  \bibinfo{author}{\bibfnamefont{X.}~\bibnamefont{Wang}},
  \bibinfo{author}{\bibfnamefont{Y.}~\bibnamefont{Kim}},
  \bibinfo{author}{\bibfnamefont{T.~C.} \bibnamefont{Shyu}},
  \bibinfo{author}{\bibfnamefont{J.}~\bibnamefont{Lyu}}, \bibnamefont{and}
  \bibinfo{author}{\bibfnamefont{N.~A.} \bibnamefont{Kotov}},
  \bibinfo{journal}{ACS nano} \textbf{\bibinfo{volume}{10}},
  \bibinfo{pages}{6156} (\bibinfo{year}{2016}).

\bibitem[{\citenamefont{Rogers et~al.}(2016)\citenamefont{Rogers, Huang,
  Schmidt, and Gracias}}]{Rogers2016}
\bibinfo{author}{\bibfnamefont{J.}~\bibnamefont{Rogers}},
  \bibinfo{author}{\bibfnamefont{Y.}~\bibnamefont{Huang}},
  \bibinfo{author}{\bibfnamefont{O.~G.} \bibnamefont{Schmidt}},
  \bibnamefont{and} \bibinfo{author}{\bibfnamefont{D.~H.}
  \bibnamefont{Gracias}}, \bibinfo{journal}{Mrs Bulletin}
  \textbf{\bibinfo{volume}{41}}, \bibinfo{pages}{123} (\bibinfo{year}{2016}).

\bibitem[{\citenamefont{Baldwin and Meng}(2017)}]{Baldwin2017}
\bibinfo{author}{\bibfnamefont{A.}~\bibnamefont{Baldwin}} \bibnamefont{and}
  \bibinfo{author}{\bibfnamefont{E.}~\bibnamefont{Meng}}, in
  \emph{\bibinfo{booktitle}{Micro Electro Mechanical Systems (MEMS), 2017 IEEE
  30th International Conference on}} (\bibinfo{organization}{IEEE},
  \bibinfo{year}{2017}), pp. \bibinfo{pages}{227--230}.

\bibitem[{\citenamefont{Saito et~al.}(2011)\citenamefont{Saito, Agnese, and
  Scarpa}}]{Saito2011}
\bibinfo{author}{\bibfnamefont{K.}~\bibnamefont{Saito}},
  \bibinfo{author}{\bibfnamefont{F.}~\bibnamefont{Agnese}}, \bibnamefont{and}
  \bibinfo{author}{\bibfnamefont{F.}~\bibnamefont{Scarpa}},
  \bibinfo{journal}{Journal of intelligent material systems and structures}
  \textbf{\bibinfo{volume}{22}}, \bibinfo{pages}{935} (\bibinfo{year}{2011}).

\bibitem[{\citenamefont{Sareh and Rossiter}(2012)}]{Sareh2012}
\bibinfo{author}{\bibfnamefont{S.}~\bibnamefont{Sareh}} \bibnamefont{and}
  \bibinfo{author}{\bibfnamefont{J.}~\bibnamefont{Rossiter}},
  \bibinfo{journal}{Smart Materials and Structures}
  \textbf{\bibinfo{volume}{22}}, \bibinfo{pages}{014004}
  (\bibinfo{year}{2012}).

\bibitem[{\citenamefont{Zhang and Dai}(2014)}]{Zhang2014}
\bibinfo{author}{\bibfnamefont{K.}~\bibnamefont{Zhang}} \bibnamefont{and}
  \bibinfo{author}{\bibfnamefont{J.~S.} \bibnamefont{Dai}},
  \bibinfo{journal}{Journal of Mechanisms and Robotics}
  \textbf{\bibinfo{volume}{6}}, \bibinfo{pages}{021007} (\bibinfo{year}{2014}).

\bibitem[{\citenamefont{Lamoureux et~al.}(2015)\citenamefont{Lamoureux, Lee,
  Shlian, Forrest, and Shtein}}]{Lamoureux2015}
\bibinfo{author}{\bibfnamefont{A.}~\bibnamefont{Lamoureux}},
  \bibinfo{author}{\bibfnamefont{K.}~\bibnamefont{Lee}},
  \bibinfo{author}{\bibfnamefont{M.}~\bibnamefont{Shlian}},
  \bibinfo{author}{\bibfnamefont{S.~R.} \bibnamefont{Forrest}},
  \bibnamefont{and} \bibinfo{author}{\bibfnamefont{M.}~\bibnamefont{Shtein}},
  \bibinfo{journal}{Nature Communications} \textbf{\bibinfo{volume}{6}},
  \bibinfo{pages}{8092} (\bibinfo{year}{2015}).

\bibitem[{\citenamefont{Hui et~al.}(1998)\citenamefont{Hui, Zehnder, and
  Potdar}}]{Hui1998}
\bibinfo{author}{\bibfnamefont{C.~Y.} \bibnamefont{Hui}},
  \bibinfo{author}{\bibfnamefont{A.~T.} \bibnamefont{Zehnder}},
  \bibnamefont{and} \bibinfo{author}{\bibfnamefont{Y.~K.}
  \bibnamefont{Potdar}}, \bibinfo{journal}{International Journal of Fracture}
  \textbf{\bibinfo{volume}{93}}, \bibinfo{pages}{409} (\bibinfo{year}{1998}).

\bibitem[{\citenamefont{Brighenti}(2005)}]{Brighenti2005}
\bibinfo{author}{\bibfnamefont{R.}~\bibnamefont{Brighenti}},
  \bibinfo{journal}{Thin-Walled Structures} \textbf{\bibinfo{volume}{43}},
  \bibinfo{pages}{209} (\bibinfo{year}{2005}).

\bibitem[{\citenamefont{Zehnder and Viz}(2005)}]{Zehnder2005}
\bibinfo{author}{\bibfnamefont{A.~T.} \bibnamefont{Zehnder}} \bibnamefont{and}
  \bibinfo{author}{\bibfnamefont{M.~J.} \bibnamefont{Viz}},
  \bibinfo{journal}{Applied Mechanics Reviews} \textbf{\bibinfo{volume}{58}},
  \bibinfo{pages}{37} (\bibinfo{year}{2005}).

\bibitem[{\citenamefont{Adda-Bedia and Ben~Amar}(2001)}]{Adda2001}
\bibinfo{author}{\bibfnamefont{M.}~\bibnamefont{Adda-Bedia}} \bibnamefont{and}
  \bibinfo{author}{\bibfnamefont{M.}~\bibnamefont{Ben~Amar}},
  \bibinfo{journal}{Physical Review Letters} \textbf{\bibinfo{volume}{86}},
  \bibinfo{pages}{5703} (\bibinfo{year}{2001}).

\bibitem[{\citenamefont{Bazant and Cedolin}(2010)}]{Bazant2010}
\bibinfo{author}{\bibfnamefont{Z.~P.} \bibnamefont{Bazant}} \bibnamefont{and}
  \bibinfo{author}{\bibfnamefont{L.}~\bibnamefont{Cedolin}},
  \emph{\bibinfo{title}{Stability of Structures: Elastic, Inelastic, Fracture
  and Damage Theories}} (\bibinfo{publisher}{World Scientific},
  \bibinfo{year}{2010}), ISBN \bibinfo{isbn}{9789814317030}.

\bibitem[{\citenamefont{Muskhelishvili}(2013)}]{Muskhelishvili2013}
\bibinfo{author}{\bibfnamefont{N.~I.} \bibnamefont{Muskhelishvili}},
  \emph{\bibinfo{title}{Some Basic Problems of the Mathematical Theory of
  Elasticity}} (\bibinfo{publisher}{Springer Netherlands},
  \bibinfo{year}{2013}), ISBN \bibinfo{isbn}{9789401730341}.

\bibitem[{\citenamefont{Jiang et~al.}(2013)\citenamefont{Jiang, Qi, Park, and
  Rabczuk}}]{jiang-Nanotechnology-24-435705-2013}
\bibinfo{author}{\bibfnamefont{J.-W.} \bibnamefont{Jiang}},
  \bibinfo{author}{\bibfnamefont{Z.}~\bibnamefont{Qi}},
  \bibinfo{author}{\bibfnamefont{H.~S.} \bibnamefont{Park}}, \bibnamefont{and}
  \bibinfo{author}{\bibfnamefont{T.}~\bibnamefont{Rabczuk}},
  \bibinfo{journal}{Nanotechnology} \textbf{\bibinfo{volume}{24}},
  \bibinfo{pages}{435705} (\bibinfo{year}{2013}).

\bibitem[{\citenamefont{Grosso and Mele}(2015)}]{Grosso2015}
\bibinfo{author}{\bibfnamefont{B.~F.} \bibnamefont{Grosso}} \bibnamefont{and}
  \bibinfo{author}{\bibfnamefont{E.~J.} \bibnamefont{Mele}},
  \bibinfo{journal}{Phys. Rev. Lett.} \textbf{\bibinfo{volume}{115}},
  \bibinfo{pages}{195501} (\bibinfo{year}{2015}).

\bibitem[{\citenamefont{Cortet et~al.}(2008)\citenamefont{Cortet, Huillard,
  Vanel, and Ciliberto}}]{Cortet2008}
\bibinfo{author}{\bibfnamefont{P.-P.} \bibnamefont{Cortet}},
  \bibinfo{author}{\bibfnamefont{G.}~\bibnamefont{Huillard}},
  \bibinfo{author}{\bibfnamefont{L.}~\bibnamefont{Vanel}}, \bibnamefont{and}
  \bibinfo{author}{\bibfnamefont{S.}~\bibnamefont{Ciliberto}},
  \bibinfo{journal}{Journal of Statistical Mechanics: Theory and Experiment}
  \textbf{\bibinfo{volume}{2008}}, \bibinfo{pages}{P10022}
  (\bibinfo{year}{2008}).

\bibitem[{\citenamefont{Dalbe et~al.}(2015)\citenamefont{Dalbe, Koivisto,
  Vanel, Miksic, Ramos, Alava, and Santucci}}]{Dalbe2015}
\bibinfo{author}{\bibfnamefont{M.-J.} \bibnamefont{Dalbe}},
  \bibinfo{author}{\bibfnamefont{J.}~\bibnamefont{Koivisto}},
  \bibinfo{author}{\bibfnamefont{L.}~\bibnamefont{Vanel}},
  \bibinfo{author}{\bibfnamefont{A.}~\bibnamefont{Miksic}},
  \bibinfo{author}{\bibfnamefont{O.}~\bibnamefont{Ramos}},
  \bibinfo{author}{\bibfnamefont{M.}~\bibnamefont{Alava}}, \bibnamefont{and}
  \bibinfo{author}{\bibfnamefont{S.}~\bibnamefont{Santucci}},
  \bibinfo{journal}{Physical review letters} \textbf{\bibinfo{volume}{114}},
  \bibinfo{pages}{205501} (\bibinfo{year}{2015}).

\bibitem[{\citenamefont{Bahamon et~al.}(2016)\citenamefont{Bahamon, Qi, Park,
  Pereira, and Campbell}}]{Bahamon2016}
\bibinfo{author}{\bibfnamefont{D.~A.} \bibnamefont{Bahamon}},
  \bibinfo{author}{\bibfnamefont{Z.}~\bibnamefont{Qi}},
  \bibinfo{author}{\bibfnamefont{H.~S.} \bibnamefont{Park}},
  \bibinfo{author}{\bibfnamefont{V.~M.} \bibnamefont{Pereira}},
  \bibnamefont{and} \bibinfo{author}{\bibfnamefont{D.~K.}
  \bibnamefont{Campbell}}, \bibinfo{journal}{Phys. Rev. B}
  \textbf{\bibinfo{volume}{93}}, \bibinfo{pages}{235408}
  (\bibinfo{year}{2016}),
  \urlprefix\url{https://link.aps.org/doi/10.1103/PhysRevB.93.235408}.

\bibitem[{\citenamefont{Castellanos-Gomez
  et~al.}(2013)\citenamefont{Castellanos-Gomez, Rold{\'a}n, Cappelluti,
  Buscema, Guinea, van~der Zant, and Steele}}]{Castellanos2013}
\bibinfo{author}{\bibfnamefont{A.}~\bibnamefont{Castellanos-Gomez}},
  \bibinfo{author}{\bibfnamefont{R.}~\bibnamefont{Rold{\'a}n}},
  \bibinfo{author}{\bibfnamefont{E.}~\bibnamefont{Cappelluti}},
  \bibinfo{author}{\bibfnamefont{M.}~\bibnamefont{Buscema}},
  \bibinfo{author}{\bibfnamefont{F.}~\bibnamefont{Guinea}},
  \bibinfo{author}{\bibfnamefont{H.~S.} \bibnamefont{van~der Zant}},
  \bibnamefont{and} \bibinfo{author}{\bibfnamefont{G.~A.}
  \bibnamefont{Steele}}, \bibinfo{journal}{Nano letters}
  \textbf{\bibinfo{volume}{13}}, \bibinfo{pages}{5361} (\bibinfo{year}{2013}).

\bibitem[{\citenamefont{Tsai et~al.}(2015)\citenamefont{Tsai, Tarasov, Hesabi,
  Taghinejad, Campbell, Joiner, Adibi, and Vogel}}]{Tsai2015}
\bibinfo{author}{\bibfnamefont{M.-Y.} \bibnamefont{Tsai}},
  \bibinfo{author}{\bibfnamefont{A.}~\bibnamefont{Tarasov}},
  \bibinfo{author}{\bibfnamefont{Z.~R.} \bibnamefont{Hesabi}},
  \bibinfo{author}{\bibfnamefont{H.}~\bibnamefont{Taghinejad}},
  \bibinfo{author}{\bibfnamefont{P.~M.} \bibnamefont{Campbell}},
  \bibinfo{author}{\bibfnamefont{C.~A.} \bibnamefont{Joiner}},
  \bibinfo{author}{\bibfnamefont{A.}~\bibnamefont{Adibi}}, \bibnamefont{and}
  \bibinfo{author}{\bibfnamefont{E.~M.} \bibnamefont{Vogel}},
  \bibinfo{journal}{ACS applied materials \& interfaces}
  \textbf{\bibinfo{volume}{7}}, \bibinfo{pages}{12850} (\bibinfo{year}{2015}).

\bibitem[{\citenamefont{He et~al.}(2013)\citenamefont{He, Poole, Mak, and
  Shan}}]{He2013}
\bibinfo{author}{\bibfnamefont{K.}~\bibnamefont{He}},
  \bibinfo{author}{\bibfnamefont{C.}~\bibnamefont{Poole}},
  \bibinfo{author}{\bibfnamefont{K.~F.} \bibnamefont{Mak}}, \bibnamefont{and}
  \bibinfo{author}{\bibfnamefont{J.}~\bibnamefont{Shan}},
  \bibinfo{journal}{Nano letters} \textbf{\bibinfo{volume}{13}},
  \bibinfo{pages}{2931} (\bibinfo{year}{2013}).

\bibitem[{\citenamefont{Lloyd et~al.}(2016)\citenamefont{Lloyd, Liu,
  Christopher, Cantley, Wadehra, Kim, Goldberg, Swan, and Bunch}}]{Lloyd2016}
\bibinfo{author}{\bibfnamefont{D.}~\bibnamefont{Lloyd}},
  \bibinfo{author}{\bibfnamefont{X.}~\bibnamefont{Liu}},
  \bibinfo{author}{\bibfnamefont{J.~W.} \bibnamefont{Christopher}},
  \bibinfo{author}{\bibfnamefont{L.}~\bibnamefont{Cantley}},
  \bibinfo{author}{\bibfnamefont{A.}~\bibnamefont{Wadehra}},
  \bibinfo{author}{\bibfnamefont{B.~L.} \bibnamefont{Kim}},
  \bibinfo{author}{\bibfnamefont{B.~B.} \bibnamefont{Goldberg}},
  \bibinfo{author}{\bibfnamefont{A.~K.} \bibnamefont{Swan}}, \bibnamefont{and}
  \bibinfo{author}{\bibfnamefont{J.~S.} \bibnamefont{Bunch}},
  \bibinfo{journal}{Nano letters} \textbf{\bibinfo{volume}{16}},
  \bibinfo{pages}{5836} (\bibinfo{year}{2016}).

\bibitem[{\citenamefont{Feng et~al.}(2012)\citenamefont{Feng, Qian, Huang, and
  Li}}]{Feng2012}
\bibinfo{author}{\bibfnamefont{J.}~\bibnamefont{Feng}},
  \bibinfo{author}{\bibfnamefont{X.}~\bibnamefont{Qian}},
  \bibinfo{author}{\bibfnamefont{C.-W.} \bibnamefont{Huang}}, \bibnamefont{and}
  \bibinfo{author}{\bibfnamefont{J.}~\bibnamefont{Li}},
  \bibinfo{journal}{Nature Photonics} \textbf{\bibinfo{volume}{6}},
  \bibinfo{pages}{866} (\bibinfo{year}{2012}).

\bibitem[{\citenamefont{Conley et~al.}(2013)\citenamefont{Conley, Wang,
  Ziegler, Haglund~Jr, Pantelides, and Bolotin}}]{Conley2013}
\bibinfo{author}{\bibfnamefont{H.~J.} \bibnamefont{Conley}},
  \bibinfo{author}{\bibfnamefont{B.}~\bibnamefont{Wang}},
  \bibinfo{author}{\bibfnamefont{J.~I.} \bibnamefont{Ziegler}},
  \bibinfo{author}{\bibfnamefont{R.~F.} \bibnamefont{Haglund~Jr}},
  \bibinfo{author}{\bibfnamefont{S.~T.} \bibnamefont{Pantelides}},
  \bibnamefont{and} \bibinfo{author}{\bibfnamefont{K.~I.}
  \bibnamefont{Bolotin}}, \bibinfo{journal}{Nano letters}
  \textbf{\bibinfo{volume}{13}}, \bibinfo{pages}{3626} (\bibinfo{year}{2013}).

\bibitem[{\citenamefont{Johari and Shenoy}(2012)}]{Johari2012}
\bibinfo{author}{\bibfnamefont{P.}~\bibnamefont{Johari}} \bibnamefont{and}
  \bibinfo{author}{\bibfnamefont{V.~B.} \bibnamefont{Shenoy}},
  \bibinfo{journal}{ACS nano} \textbf{\bibinfo{volume}{6}},
  \bibinfo{pages}{5449} (\bibinfo{year}{2012}).

\bibitem[{\citenamefont{Kotov et~al.}(2012)\citenamefont{Kotov, Uchoa, Pereira,
  Guinea, and Neto}}]{Kotov2012}
\bibinfo{author}{\bibfnamefont{V.~N.} \bibnamefont{Kotov}},
  \bibinfo{author}{\bibfnamefont{B.}~\bibnamefont{Uchoa}},
  \bibinfo{author}{\bibfnamefont{V.~M.} \bibnamefont{Pereira}},
  \bibinfo{author}{\bibfnamefont{F.}~\bibnamefont{Guinea}}, \bibnamefont{and}
  \bibinfo{author}{\bibfnamefont{A.~C.} \bibnamefont{Neto}},
  \bibinfo{journal}{Reviews of Modern Physics} \textbf{\bibinfo{volume}{84}},
  \bibinfo{pages}{1067} (\bibinfo{year}{2012}).

\bibitem[{\citenamefont{COMSOL}(Last accessed 14 July 2017)}]{COMSOL}
\bibinfo{author}{\bibnamefont{COMSOL}},
  \bibinfo{journal}{http://www.comsol.com/comsol-multiphysics}
  (\bibinfo{year}{Last accessed 14 July 2017}).

\bibitem[{\citenamefont{Saha and Culpepper}(2014)}]{saha2014meshperturb}
\bibinfo{author}{\bibfnamefont{S.~K.} \bibnamefont{Saha}} \bibnamefont{and}
  \bibinfo{author}{\bibfnamefont{M.~L.} \bibnamefont{Culpepper}}
  (\bibinfo{year}{2014}).

\bibitem[{\citenamefont{Lammps}(2012)}]{plimptonLAMMPS}
\bibinfo{author}{\bibnamefont{Lammps}},
  \bibinfo{journal}{http://lammps.sandia.gov}  (\bibinfo{year}{2012}).

\bibitem[{\citenamefont{Stuart et~al.}(2000)\citenamefont{Stuart, Tutein, and
  Harrison}}]{stuart2000reactive}
\bibinfo{author}{\bibfnamefont{S.~J.} \bibnamefont{Stuart}},
  \bibinfo{author}{\bibfnamefont{A.~B.} \bibnamefont{Tutein}},
  \bibnamefont{and} \bibinfo{author}{\bibfnamefont{J.~A.}
  \bibnamefont{Harrison}}, \bibinfo{journal}{The Journal of chemical physics}
  \textbf{\bibinfo{volume}{112}}, \bibinfo{pages}{6472} (\bibinfo{year}{2000}).

\bibitem[{\citenamefont{Jiang}(2015)}]{jiang-Nanotechnology-26-315706-2015}
\bibinfo{author}{\bibfnamefont{J.-W.} \bibnamefont{Jiang}},
  \bibinfo{journal}{Nanotechnology} \textbf{\bibinfo{volume}{26}},
  \bibinfo{pages}{315706} (\bibinfo{year}{2015}).

\bibitem[{\citenamefont{Hanakata et~al.}(2016)\citenamefont{Hanakata, Qi,
  Campbell, and Park}}]{hanakata-Nanoscale-8-458-2016}
\bibinfo{author}{\bibfnamefont{P.~Z.} \bibnamefont{Hanakata}},
  \bibinfo{author}{\bibfnamefont{Z.}~\bibnamefont{Qi}},
  \bibinfo{author}{\bibfnamefont{D.~K.} \bibnamefont{Campbell}},
  \bibnamefont{and} \bibinfo{author}{\bibfnamefont{H.~S.} \bibnamefont{Park}},
  \bibinfo{journal}{Nanoscale} \textbf{\bibinfo{volume}{8}},
  \bibinfo{pages}{458} (\bibinfo{year}{2016}).

\end{thebibliography}

\end{document}